# Skyrmions based on optical anisotropy for topological encoding

Yunqi Zhang[†], An Aloysius Wang[†], Runchen Zhang[†], Zimo Zhao, Yifei Ma, Ruofu Liu, Zhi-Kai Pong, Yuxi Cai, and Chao He[*]

*Department of Engineering Science, University of Oxford, Parks Road, Oxford, OX1 3PJ, UK*
*†These authors contributed equally to this work*
*[*]Corresponding author: chao.he@eng.ox.ac.uk*

## Abstract

**The observation of skyrmions across diverse physical domains suggests that they are universal features of $S^2$-valued fields, reflecting the ubiquity of topology in the study of the natural world. In this paper, we develop an abstract technique of parameter space dimensionality reduction that extends the skyrmion framework to fields taking values in manifolds of dimension greater than 2, thereby broadening the range of systems that can support skyrmions. To prove that this is more than just a mathematical abstraction, we apply our technique to light-matter interactions, directly encoding skyrmionic structures into the optical anisotropy of spatially varying structured matter by selecting a distinguished axis, an approach fundamentally different from the more commonly known skyrmions formed by director fields in liquid crystals. We experimentally realize such skyrmions using a liquid-crystal-based tunable elliptical retarder array as a proof-of-concept platform and demonstrate complex, reconfigurable skyrmionic states exhibiting topological robustness under artificially introduced stochastic perturbations. Exploiting this robustness, we demonstrate a promising application of skyrmions in topologically protected information storage and provide an easily verifiable rule, which we term the 60° rule, that serves as a practical engineering criterion for guaranteeing robustness against noise and measurement errors.**

Skyrmions are topologically nontrivial field configurations that were originally proposed in particle physics[1,2] and have since been realized in a range of naturally $S^2$-valued physical systems, most notably in magnetic spin textures[3–9]. Recently, there has been a growing interest in extending the framework of skyrmions to more unconventional systems that are vector-valued but not necessarily $S^2$-valued. This includes recent work in water waves[10], acoustic systems[11,12], optical and photonic platforms[13–20], and more[21–25]. To interpret such fields as skyrmions, it is often implicitly assumed that the physically relevant field is projected – typically through normalization – onto the 2-sphere, after which standard considerations can be applied. Given this interpretation, we propose an entirely new generalization to higher dimensional fields, whereby different physically relevant $S^2$-valued subcomponents of the field are used to define skyrmions[26].

From a practical perspective, this technique of *parameter space dimensionality reduction* opens the door to engineering high-dimensional fields that carry topological information, with the potential for a single field to encode multiple topological charges through different projection maps. This perspective is particularly significant for robust information encoding, one of the key applications of skyrmions, and is supported by recent work showing that topological properties can emerge from appropriately constructed mappings of high-dimensional parameter spaces rather than being exclusively tied to real-space field textures[27]. The added flexibility and information density enabled by parameter space reduction may therefore prove invaluable for real-world applications.

More concretely, various candidates for skyrmion-based information encoding have already been proposed[28–30], including magnetic spin textures, optical fields, and the director fields of liquid crystals (LCs). However, each of these strategies presents some limitations. Magnetic skyrmions, while compact, are generally temperature sensitive[31–33]. Optical skyrmions, although easily generated, are transient in nature and thus more suitable for information transfer than storage[34]. Lastly, LCs are limited to low orders due to energy considerations[35] and the stored topological charge cannot be optically read out easily[36]. Moreover, the nonpolar nature of LCs[37,38] means that it is impossible to distinguish between positive and negative skyrmion numbers, limiting information density.

In this paper, we use the flexibility of parameter space dimensionality reduction to propose a strategy for topological information encoding that overcomes the limitations of magnetic, optical and LC skyrmions mentioned above. This is achieved not by considering matter fields or light fields alone, but rather by focusing on the light-matter interactions of complex, spatially varying structured matter[39]. Structured matter is an umbrella term for various physical media that manipulate the vectorial properties of light and are often described by a spatially varying Jones or Mueller matrix, depending on whether the material is depolarizing. Since the Jones matrix field is described by a spatially varying $2 \times 2$ complex matrix, while the Mueller matrix field is described by a spatially varying $4 \times 4$ real matrix, both are high dimensional (8 and 16, respectively) and can support many non-trivial $S^2$-valued subcomponents. This suggests that multiple, potentially distinct topological numbers can arise via parameter-space dimensionality reduction. In particular, many physically meaningful maps $\mathcal{P}$ can be devised to highlight the topology of different aspects of the underlying field. One representative example is provided by the Mueller matrix polar decomposition, in which a Mueller matrix is decomposed into a depolarizer, a diattenuator, and a retarder[40]. These components provide additional degrees of freedom for engineering skyrmion structures with novel physical properties, as detailed in the Results.

Here, we focus on one such projection map, termed $\mathcal{P}_{\text{aniso}}$, which picks out a distinguished axis in materials exhibiting optical anisotropy. Not only are such skyrmions directly optically readable, but they can also encode information permanently, making them practical candidates for optical data storage applications. In this work, we use a spatially varying elliptical retarder array[39] as a preliminary platform for demonstration, with the construction of the map $\mathcal{P}_{\text{aniso}}$ detailed in Methods 1.

Lastly, to demonstrate the capabilities of our approach, we first use the tunable anisotropy of the retarder array to freely manipulate the effective distinguished axis of the entire cascade[41,42] and generate a range of complex skyrmionic field configurations — including higher-order skyrmions, skyrmion bags, skyrmion lattices, and meron lattices. We then assess topological robustness by introducing varying levels of perturbation to a representative skyrmion (Néel-type skyrmion), confirming its stability under noise. This is in line with a newly established sufficient condition for topological protection, referred to as the 60° rule, which serves as a practical and easily verifiable engineering criterion for guaranteeing robustness in the presence of noise and measurement errors. The rule states that the encoded information remains topologically protected provided the physically realized $S^2$-valued field differs from the designed field everywhere by up to 60°, which is a remarkable margin of error[18] (see Methods 2 and Supplementary Note 4). Finally, we demonstrate a proof-of-concept optical encoding scheme by storing ASCII information in perturbed skyrmion bags, highlighting their potential for robust optical data storage. These findings establish a new paradigm for topological data storage, providing a foundation for next-generation programmable information storage and processing that is both high-density and perturbation resistant.

## Results

We begin by describing *parameter space dimensionality reduction* and giving several examples which demonstrate its utility. Given a smooth field $f$ taking values in a manifold $M$ of dimension greater than 2, one can construct an $S^2$-valued field through a surjective smooth projection $\mathcal{P}: M \longrightarrow S^2$ by considering the composition $\mathcal{P} \circ f$. For example, given a spatially varying optical retarder, the corresponding Jones matrix field takes values in $M = SU(2)$. One can then take the projection map to be the standard Hopf map $\mathcal{P}_{\mathrm{Hopf}}$, which physically corresponds to factoring out retardance information. With this specific pairing of $M$ and $\mathcal{P}$, the results of Ref[43] can be rephrased as follows: a $\mathcal{P}_{\mathrm{Hopf}}$-meron of order $k$ is an optical Stokes skyrmion adder of order $k$. Thus, in this instance, the parameter space reduction perspective builds a strong link between the topological number of the underlying matter field and its behavior on light fields, further strengthening the notion of topological duality introduced in Ref[43].

More generally, we do not need $\mathcal{P}$ to be well-defined everywhere. In this case, the values where $\mathcal{P}$ is not defined correspond to singularities. In fact, many commonly used projection maps $\mathcal{P}$ are singular. For example, in water wave skyrmions, $\mathcal{P}$ acts on velocity fields by normalization, which is not well-defined at zero, whereas in optical Stokes skyrmions, $\mathcal{P}$ acts on electric fields by normalization (corresponding to factoring out intensity information) followed by the Hopf map (corresponding to factoring out phase) and is likewise not well-defined at zero. From a practical perspective, having a singular projection map opens the doors to more complex topological textures and the capacity to shape singularities within a field. Given the framework of generalized skyrmions[44], the ability to incorporate and manipulate singularities within an $S^2$-valued field has the potential to exponentially increase topological information density. We leave this remark as a possible extension of the work presented here.

Turning to our main focus, we consider spatially varying optical retarders described by Mueller matrices. In this case, $M$ is a submanifold of $\mathbb{R}^{16}$ which is isomorphic to $SO(3)$. The projection map $\mathcal{P}_{\mathrm{aniso}}$ then picks out the axis of rotation, corresponding to the fast axis of the optical retarder (see Methods 1). Not only is the proposed map physically meaningful, but it can also be generalized to more complex systems with diattenuation and depolarization by picking out the rotation matrix associated with the corresponding Mueller matrix polar decomposition[40]. We call the image of $\mathcal{P}_{\mathrm{aniso}}$ the axis geometry of the optical retarder and denote the corresponding field by $\boldsymbol{A}$ (see Fig. 1a) and refer to such skyrmions as axis-geometry-based (AGB) skyrmions.

Throughout this work, a spatially varying tunable elliptical retarder array is used to form our proposed optical anisotropy based skyrmions. The array is implemented using cascaded LC spatial light modulators (LC-SLMs) to form a pixel-level controllable and reconfigurable array[39] (see Fig. 1b and Ref[39]). We emphasize again that this platform is used for demonstration, and our proposed concept is readily applicable to other systems. The skyrmionic structures are decoded using Mueller matrix polarimetry[45,46], which provides complete vectorial characterization of the medium.

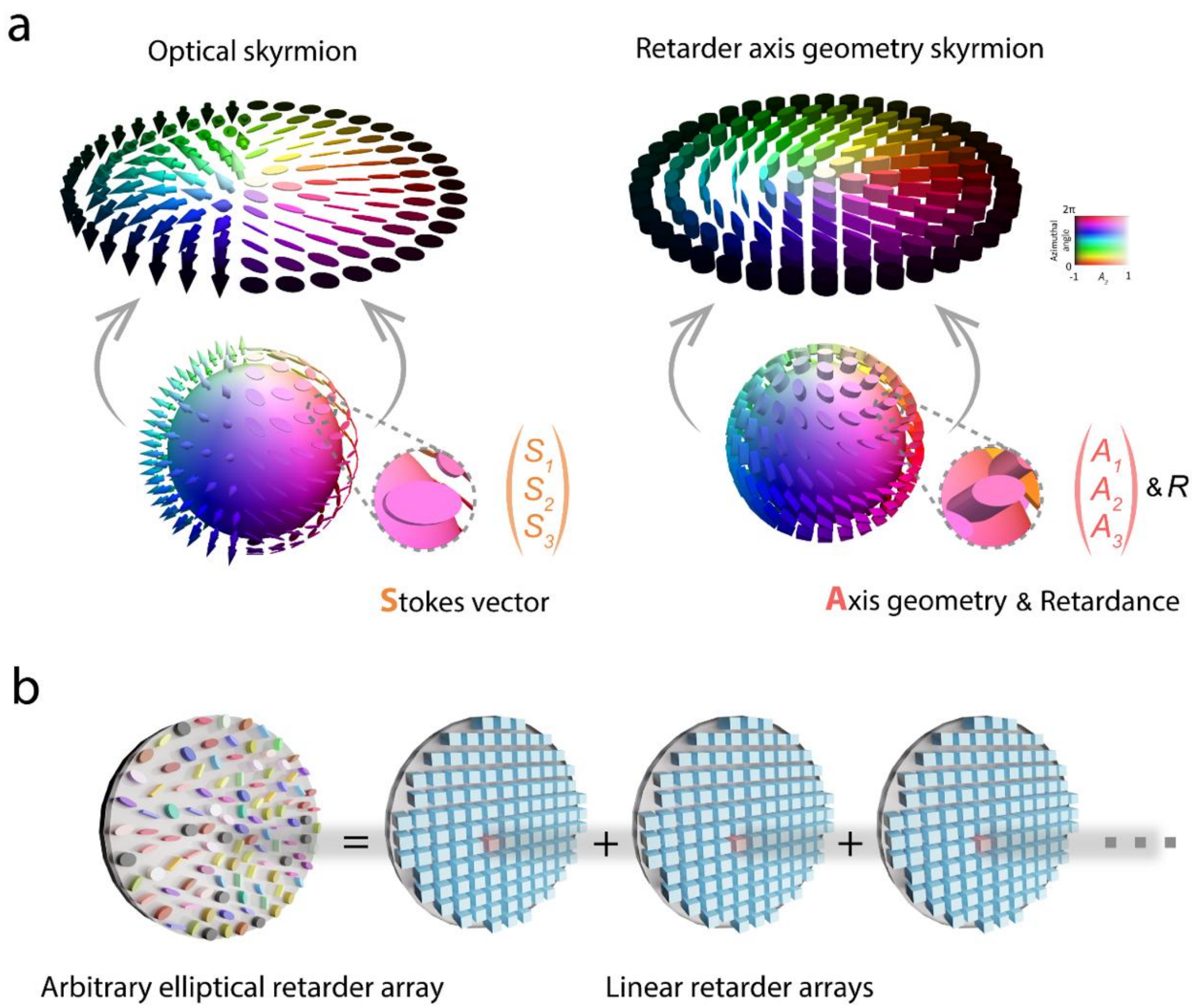

Figure 1 **Concept**. **a**. Any vector field with a parameter space of $S^2$ can be used to construct a skyrmion. (Left) A Stokes vector field $\boldsymbol{S}(x,y) = [S_1(x,y), S_2(x,y), S_3(x,y)]^T$, which is commonly depicted by arrows or polarization ellipses. (Right) An arbitrary elliptical retarder array is characterized by its axis geometry $\boldsymbol{A}$ and retardance $R$ (note induced phase and attenuation are not discussed throughout this work) and commonly depicted by elliptical cylinders, where the shape of the cylinder describes the axis geometry while the height describes the retardance. The axis geometry field of an arbitrary elliptical retarder array as defined in the main text, denoted as $\boldsymbol{A}(x,y) = [A_1(x,y), A_2(x,y), A_3(x,y)]^T$, maps the domain into $S^2$ and can thus be used to construct skyrmions. Additionally, vector fields are depicted using hue to indicate the azimuthal angle and saturation to represent $S_3$ and $A_3$. **b**. An arbitrary elliptical retarder array can be formed by cascading multiple linear retarder arrays. Each linear retarder contributes a defined retardance and fast axis orientation, which together determine the overall axis geometry and net retardance of the synthesized elliptical retarder.

To validate the feasibility of our framework, we experimentally realized various skyrmionic structures using LC-SLMs. The process involves three key steps: 1) encoding the desired spatially varying patterns into the axis geometry of the synthesized retarder array (see Supplementary Note 2); 2) measuring the corresponding Mueller matrices via polarimetry; and 3) extracting axis information through matrix decomposition[47] (see Supplementary Note 1). Using this procedure, we generate six distinct skyrmionic structures, with both theoretical and experimental results shown in Fig. 2a, and two representative Mueller matrix examples illustrated in Fig. 2b. We also calculated the skyrmion numbers for these results using the integral formula:

$$N_{\mathrm{sk}} = \frac{1}{4\pi} \iint \boldsymbol{A} \cdot \left( \frac{\partial \boldsymbol{A}}{\partial x} \times \frac{\partial \boldsymbol{A}}{\partial y} \right) dxdy \tag{1}$$

where $\boldsymbol{A}$ is the axis geometry field. The measured results closely match the theoretical predictions in terms of skyrmion numbers and structures. This agreement confirms that our method can generate various topological textures by manipulating the optical anisotropy into spatially nontrivial distributions. Furthermore, it shows that polarimetry can reliably determine the Mueller matrices without ambiguity, thereby consistently reading out the stored topological information. Note that the diversity of accessible patterns highlights the versatility of our proposed approach, whose formation is decoupled from elastic energy constraints (see Ref[48]), offering a robust platform for broader topological structure engineering.

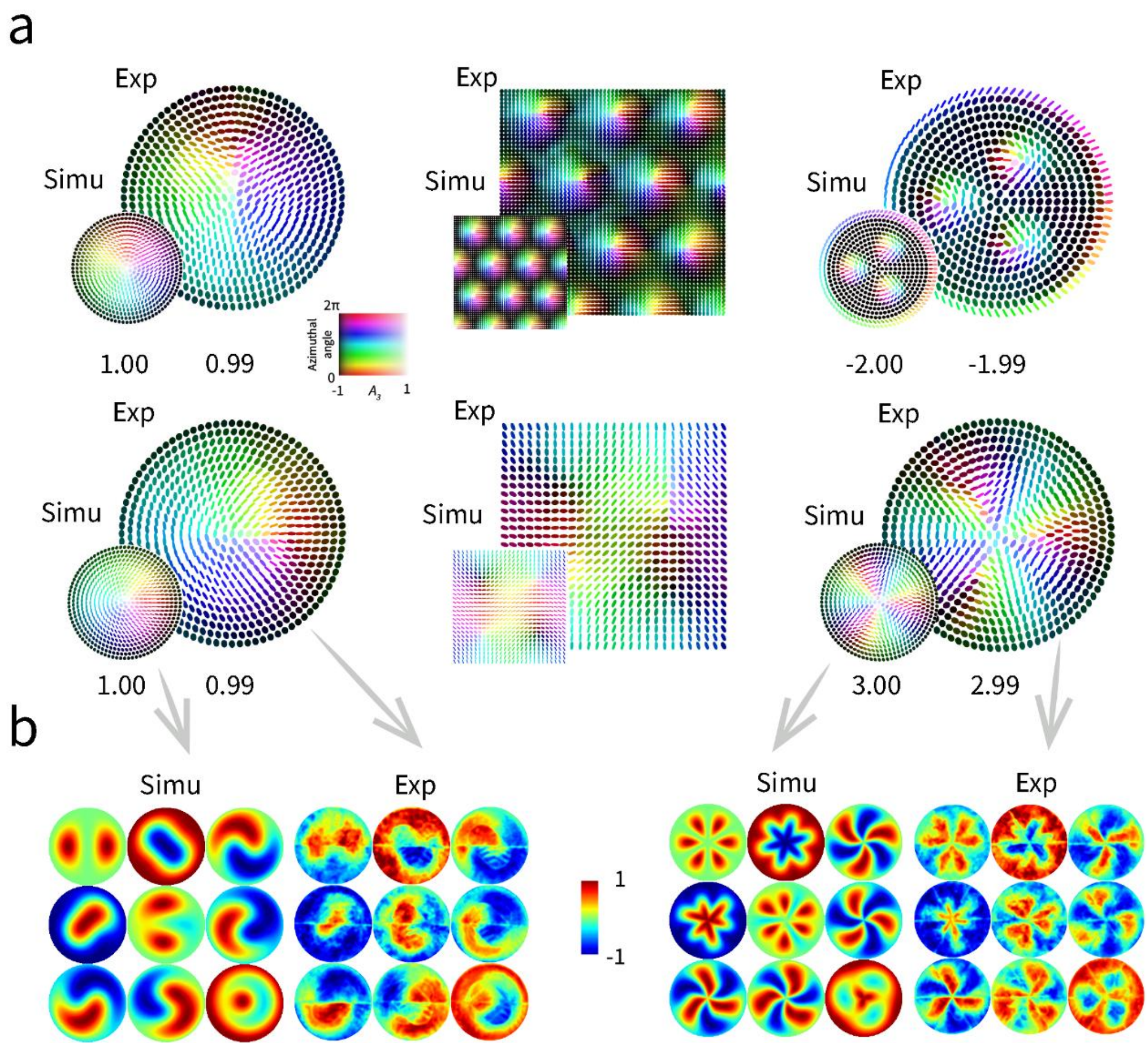

Figure 2 **Formation and characterization of optical anisotropy-based skyrmions**. **a**. Experimental (Exp) and simulated (Simu) results of the axis geometry fields of various skyrmions (for experimental setup and specifications, see Supplementary Note 2), (top row, left to right) Bloch-type skyrmion, skyrmion lattice, skyrmion bag, (bottom row, left to right) Néel-type skyrmion, meron lattice, higher-order skyrmion. The axis geometry fields are visualized using ellipses and the same color representation as introduced in Fig. 1. Calculated skyrmion numbers for experimental and simulated results are indicated below each axis geometry field plot. **b**. Mueller matrix characterization of Néel-type skyrmion (left) and higher-order skyrmion (right).

To demonstrate the topological stability of AGB skyrmions and highlight their potential for perturbation-resistant data storage, we conducted experiments under controlled Gaussian noise. While the impact of noise on skyrmion stability has been explored in structured light[43,49,50], this problem remains in its infancy. Here, we extend these results to structured matter. The procedure is as follows: 1) we first formed a unit skyrmion, which was characterized via Mueller matrix measurements and decomposition, following the same method described above; 2) external Gaussian noise was artificially added to the retardance of the different layers which make up the structured matter, with the noise standard deviation (σ) varied from 0 to π in increments of 0.05π. Note that real-world disturbances such as thermal drift and mechanical vibrations — which are inherently non-Gaussian — were also present during the experiments; 3) to quantify the effects of perturbation, we performed 120 independently generated noise samples across all noise levels, with six repeated trials per $\sigma$ value, and computed the skyrmion number for each case, extracting the mean and standard deviation.

The results, summarized in Fig. 3a, reveal three stability regimes. In the low-noise regime ($\sigma \leq 0.3\pi$), the skyrmion number remains stable at $N_{sk} \approx 1.0$ with virtually no deviation between different noise samples, confirming the strong topological protection of AGB skyrmions. As $\sigma$ increases, the skyrmion number begins to oscillate, and in the high-noise regime ($\sigma \geq 0.45\pi$) it collapses completely, indicating irreversible loss of topological information. This is qualitatively consistent with our theoretical analysis, which demonstrates that topological protection holds whenever the pointwise error of the constructed field is less than 60° (Methods 2 and Supplementary Note 4). Next, representative axis field patterns taken from a random sample at each noise level are shown in Fig. 3b. The combined data from Fig. 3a (quantitative) and Fig. 3b (visual field distributions) taken together support the conclusion that AGB skyrmions retain their topological charges under moderate perturbations. This robustness — primarily governed by boundary conditions (see Supplementary Note 3 and Ref[18]) — highlights a key advantage of exploiting topology in data storage: the ability to reliably decode information even in the presence of real-world disorder.

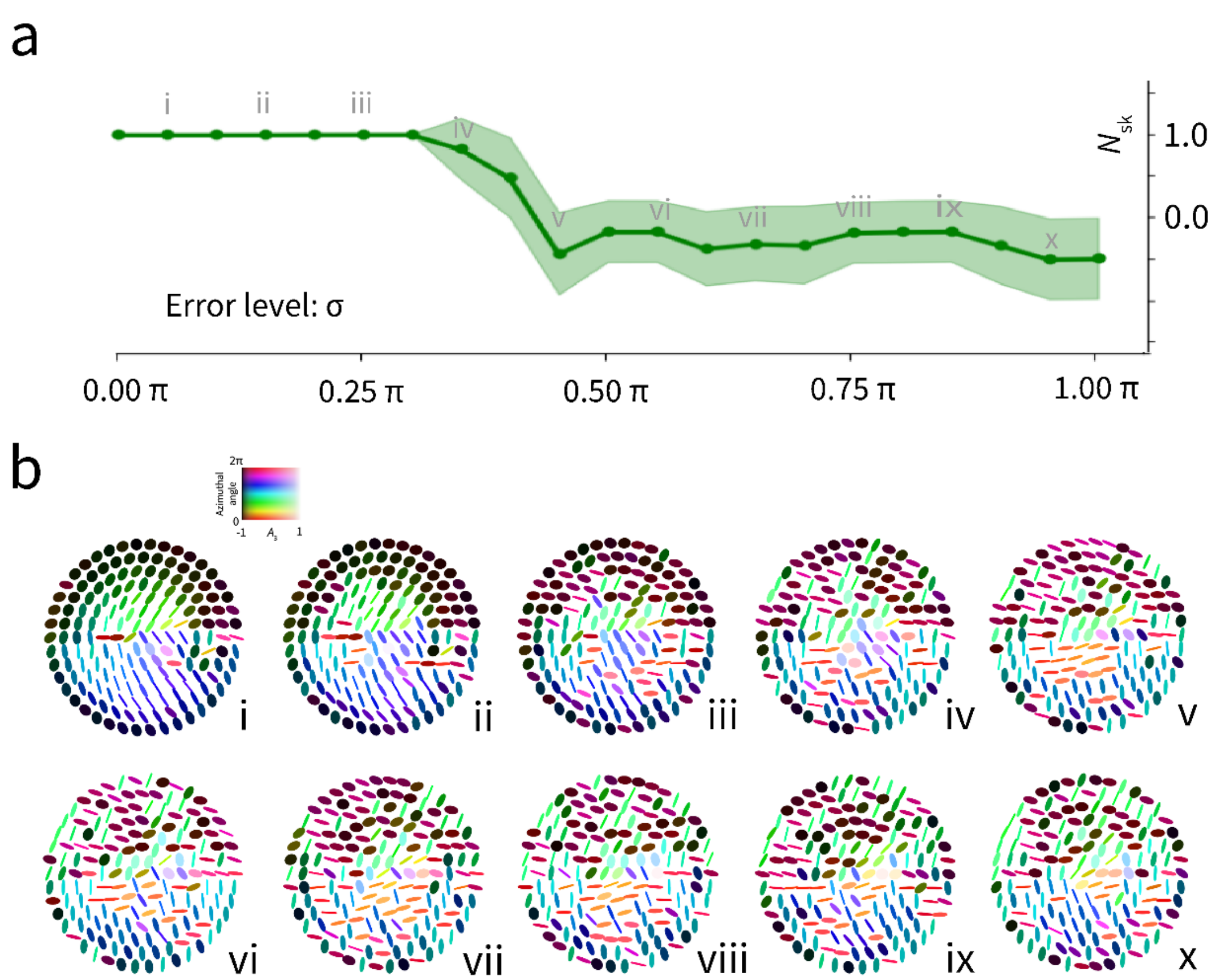

Figure 3 **Skyrmion numbers and matter-field configurations of a Néel-type AGB skyrmion under increasing perturbation**. **a.** Gaussian noise with standard deviation ($\sigma$) ranging from 0 to $\pi$ was introduced to create controlled perturbations (see details in Methods 2). At each noise level, the mean skyrmion number $N_{sk}$ (data points) and its standard deviation $\sigma_{N_{sk}}$ (shaded region) were determined through repeated

measurements. **b.** Representative AGB skyrmion patterns at ten equally spaced noise levels (i–x, corresponding to markers in panel **a**). In the low noise regime (i-iii), the internal axis field is distorted while the boundary remains intact, and the skyrmion number stays topologically protected. When the noise reaches a moderate level (iv), the skyrmion number begins to fluctuate, indicating that topological protection is no longer guaranteed to hold. Under higher noise (v–x), the boundary conditions are disrupted, resulting in the collapse of topological protection.

Lastly, to verify the application potential of AGB skyrmions in information storage, we present an example utilizing ASCII encoding to achieve reconfigurable and topologically protected storage. Our implementation is based on the structure of a skyrmion bag. The four skyrmions within the skyrmion bag encode information using skyrmion numbers ranging from $N_{\mathrm{sk}} = -2$ to $N_{\mathrm{sk}} = 2$, which enables the storage of two 16-bit numbers in a single bag. As illustrated in Fig. 4a, the encoding proceeds as follows: the chirality of the enclosing skyrmion in the skyrmion bag dictates the order in which the four enclosed skyrmions are read. By designating one enclosed skyrmion as the starting reference (in our case, the rightmost one), the four skyrmion numbers $(\alpha, \beta, \gamma, \delta)$ uniquely define two 16-bit numbers (see Supplementary Note 5 for the encoding scheme).

We experimentally encoded six letters spelling "SKYRME" and measured the skyrmion numbers of the four enclosed skyrmions in each skyrmion bag. The results in Fig. 4b show that the AGB skyrmions preserve their encoded information despite the presence of perturbations, with measured skyrmion numbers closely matching the target values. This confirms the feasibility of using perturbation-resistant AGB skyrmions for robust data storage. Importantly, the memory strategy is not limited to skyrmion bags and provides a generalized framework applicable to a wide range of topological configurations. For example, structured tailoring of skyrmion lattices or bi-skyrmions may further enhance information density.

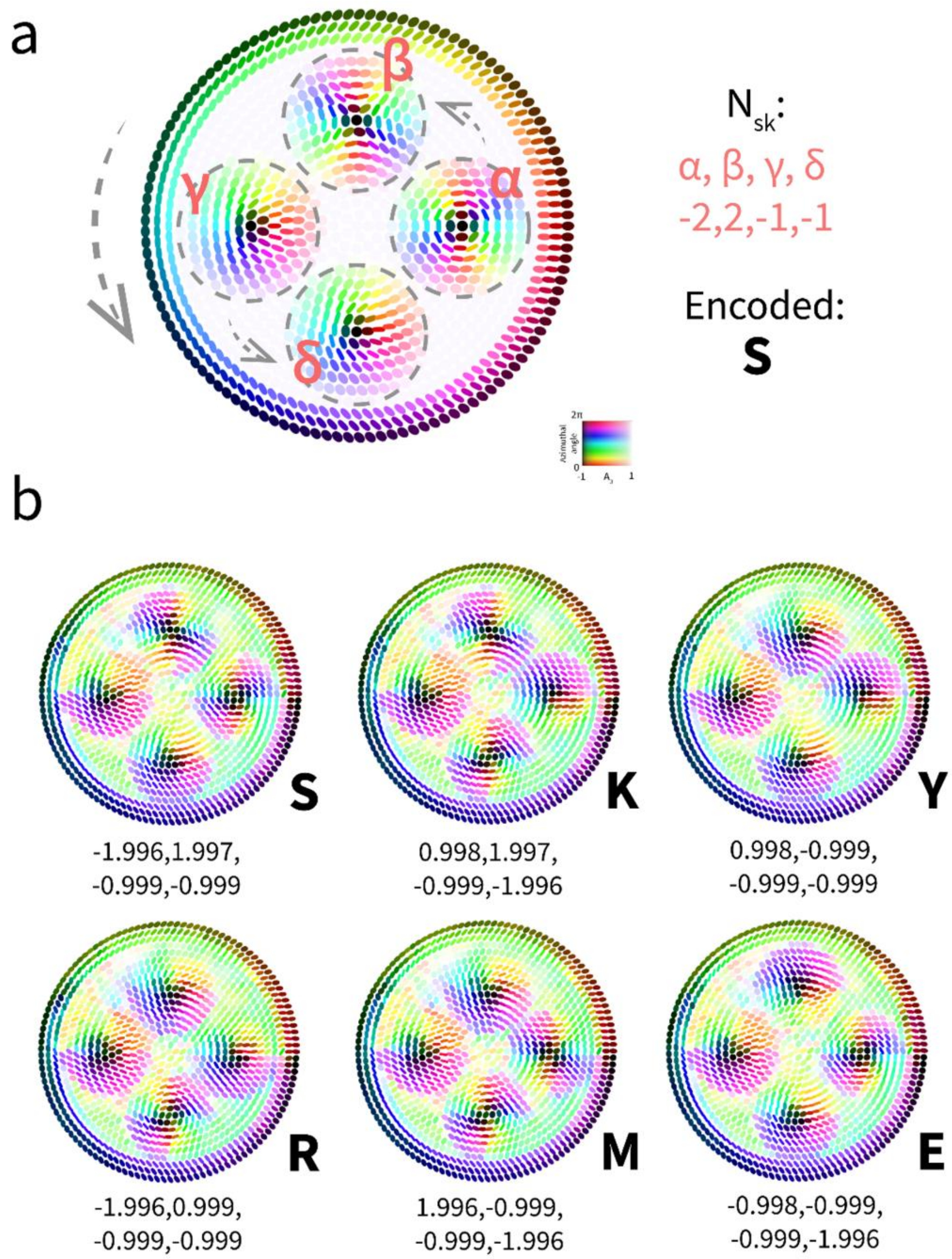

Figure 4 **AGB skyrmion bags for topological information encoding**. **a**. Example of encoding information within a single skyrmion bag containing four enclosed skyrmions, each with a skyrmion number ranging from −2 to 2. The chirality of the enclosing skyrmion defines the reading direction for the internal skyrmion numbers. Starting from the rightmost skyrmion as $\alpha$, the values $(\alpha, \beta, \gamma, \delta)$ are read sequentially along the direction indicated by gray arrows. These four values form two pairs, each encoding one 16-bit number, thereby enabling ASCII character encoding (see Supplementary Note 5). In this example, the encoded character is "S". **b**. Experimentally designed and measured skyrmion bags encoding the six characters "SKYRME". For each bag, the four enclosed skyrmion numbers are extracted and mapped to ASCII characters based on the encoding rule shown in panel **a,** and further detailed in Supplementary Note 5. All measurements were performed under real-world noise, demonstrating robust information retrieval under perturbations.

## Discussions

In this work, we formalize a technique of parameter space dimensionality reduction that allows higher-dimensional vector fields to be interpreted through the lens of skyrmions. This technique is specialized to the light-matter interactions of structured matter, where we propose a new class of skyrmions constructed based on optical anisotropy. These skyrmions, which we term AGB skyrmions, are then realized using LC retarder platforms, and their robustness and capacity for information storage are demonstrated experimentally.

Conceptually, the work presented here enriches the study of structured matter, which is typically treated as a means to manipulate optical fields, by assigning topological degrees to the matter-field itself. Moreover, compared with conventional LC skyrmions, our approach addresses key limitations in information storage, including the nonpolar nature of LC director fields, which constrains data density, the lack of effective decoding methods, and spatial confinement, which restricts skyrmions to low-order configurations. However, we acknowledge that our current demonstration employs a compound synthetic platform rather than a single, tunable, true elliptical retarder device (see discussion below).

Several promising directions are enabled by our framework: 1) to increase storage density, exploring new LC materials and fabrication strategies — such as tunable LC droplets acting as nanoscale information carriers[51] — may prove essential; 2) to enable rewritable and adaptive encoding, photo-responsive dyes such as SD1[52,53] and other photo-responsive dyes[54,55] offer promising integration with LC platforms; 3) to improve modulation speed, fast-response materials such as blue-phase LCs[56,57] may open new opportunities for real-time AGB skyrmion manipulation, enabling significantly reduced information writing times; 4) to truly overcome spatial confinement and realize true elliptical retarder devices with pixel-level control, new LC molecules and device architectures can be further investigated; 5) to achieve greater robustness and higher information density, the generalized skyrmion numbers can be adopted[43,44].

Importantly, as noted earlier, AGB skyrmions are not restricted to LC systems and can, in principle, be realized in any optical retarder platform with precise axis control. Candidate technologies include laser-written birefringent plates[58,59], metasurfaces[60], or gradient index lens assemblies[61–64]. Moreover, we emphasize again that the concept of anisotropy-axis geometry can be extended beyond retarders to include diattenuators[39], introducing absorber-based structures into this new skyrmion family with new physical properties yet to be explored. Beyond the encoding strategy demonstrated above, AGB skyrmion lattices can also be employed for image-based data storage (see Supplementary Note 5), offering a rich design space for future developments in topological structured matter.

Overall, we have introduced a new strategy for constructing skyrmions in unconventional, high-dimensional fields. Using light-matter interactions as a case study, we define a new type of skyrmion, the AGB skyrmion, and show that our abstract formalism enables the construction of complex topological states that would otherwise be difficult to observe. Furthermore, we demonstrate a practical use of AGB skyrmions in optically readable, topologically protected information storage. We believe this work opens new avenues for exploring topological phenomena in higher-dimensional systems and for developing robust platforms for high-density information storage and processing.

## Methods

### 1 Constructing an AGB skyrmion

Here, we define the map $\mathcal{P}_{\text{aniso}}$ used in the definition of AGB skyrmions implicitly. The primary objective is to construct an $S^2$-valued field which can be easily computed from the Mueller matrix. The formation of such a vector field is extracted from the eigenvector of the retarder matrix, which is obtained through Mueller matrix decomposition.

To understand the role of the Mueller matrix in this process, we briefly recall how it operates on light's state of polarization (SoP). Light, as a classical electromagnetic wave, possesses four fundamental degrees of freedom: wavelength, amplitude, phase, and polarization. In Mueller calculus[65], light's SoP can be represented by the Stokes vector $\boldsymbol{S}$, and the Mueller matrix $M$ characterizes the effect of a matter-field on this light field. When the incident light's Stokes vector is $\boldsymbol{S}_{\text{in}}$, the output $\boldsymbol{S}_{\text{out}}$ is obtained through $\boldsymbol{S}_{\text{out}} = M\boldsymbol{S}_{\text{in}}$. The Mueller matrix effects can be categorized into depolarization, diattenuation, and retardation[66–68]. Here, we restrict our analysis to a pure retarder, and this condition holds for all the theories and experiments presented above. Consequently, from the point of view of the Mueller matrix, only the lower-right $3 \times 3$ submatrix and $m_{11}$ remain non-zero, as follows:

$$M = \begin{bmatrix} 1 & 0 & 0 & 0 \\ 0 & m_{22} & m_{23} & m_{24} \\ 0 & m_{32} & m_{33} & m_{34} \\ 0 & m_{42} & m_{43} & m_{44} \end{bmatrix} \tag{2}$$

Generally, the lower-right submatrix is a rotation matrix, representing a rotation of the incident SoP on the Poincaré sphere, and thus belongs to the conventional three-dimensional special orthogonal group $SO(3)$. It is known that a three-dimensional rotation can be specified by a rotation axis and a rotation angle, which correspond respectively to the fast axis geometry and retardance in polarization optics. The axis geometry field can be expressed in the following normalized form:

$$\boldsymbol{A}(r,\theta) = \begin{pmatrix} A_1(r,\theta) \\ A_2(r,\theta) \\ A_3(r,\theta) \end{pmatrix} \tag{3}$$

where $A_1(r,\theta)^2 + A_2(r,\theta)^2 + A_3(r,\theta)^2 = 1$, with $r$ and $\theta$ denoting the polar radius and angle, respectively. Notably, by isolating the spatially distributed fast axis geometry, we can obtain an axis geometry field that takes values in $S^2$.

After developing the technique to construct the axis geometry field (as we have done in this work), we must tailor the field to satisfy the definition of a skyrmion. Accordingly, we design the axis geometry distribution of an AGB skyrmion as follows:

$$\begin{pmatrix} A_1(r,\theta) \\ A_2(r,\theta) \\ A_3(r,\theta) \end{pmatrix} = \begin{pmatrix} \sqrt{1-f(r)^2}\cos{(n\theta)} \\ \sqrt{1-f(r)^2}\sin{(n\theta)} \\ f(r) \end{pmatrix} \tag{4}$$
$$f(r) = \cos(\pi r),\ 0 \le r \le 1$$

where $r \in [0,1]$ and $\theta \in [0,2\pi]$ represent the polar radius and angle in the polar coordinate system, respectively. The resulting skyrmion has a target skyrmion number equal to $n$. The function $f(r) = \cos{(\pi r)}$ ensures that the vector field smoothly maps the domain $U$ onto $S^2$, with the well-defined boundary conditions $f(0) = 1$ and $f(1) = -1$. When $n = 1$, the above construction describes a standard Néel-type skyrmion. More complex topological configurations can be constructed by modifying the angular dependence $n\theta$ or the radial function $f(r)$.

## 2 Topological resilience of AGB skyrmions

Here, we demonstrate the topological stability of AGB skyrmions under external perturbations, where stability here refers to the robustness of the skyrmion number. We identify three distinct stability regimes: 1) in the low-perturbation regime, the skyrmion number remains invariant, indicating strong topological protection; 2) as the perturbation strength increases, the system enters an intermediate regime where the mean skyrmion number begins to decrease due to partial boundary distortions; 3) in the high-perturbation regime, the system undergoes a breakdown, and the skyrmion number collapses, indicating a loss of topological information.

*Topological resilience of AGB skyrmions via a homotopy approach*

We first investigate the topological stability of AGB skyrmions using the homotopy formalism introduced in Ref[18]. By noting that the topological degree remains invariant under smooth homotopies, we systematically explore the robustness of these structures against continuous deformations, providing a sufficient condition for the skyrmion field to be topologically protected. We term this sufficient condition the 60° rule, which states that topological protection is ensured provided that the deviation of the perturbed axis geometry field from the designed field does not exceed 60° anywhere.

Formally, suppose $\boldsymbol{A}: U \to S^2$ is the axis geometry field. Let $\mathcal{P}: \mathbb{R}^3 \backslash \{(0,0,0)\} \to S^2$ be the projection defined by $\mathcal{P}(\boldsymbol{\xi}) = \frac{1}{\|\xi\|}\boldsymbol{\xi}$, where $\boldsymbol{\xi}$ is an arbitrary vector, and let the perturbed axis geometry field $\boldsymbol{A}'$ be written in the form $\boldsymbol{A}' = \mathcal{P}(\boldsymbol{A} + t\boldsymbol{\delta})$ for some compactifiable $\boldsymbol{\delta}$.

Consider now the homotopy $H: U \times [0,1] \longrightarrow S^2$ given by

$$H(\boldsymbol{x}, t) = \mathcal{P}\big(A(\boldsymbol{x}) + t\delta(\boldsymbol{x})\big) \tag{5}$$

To ensure that Eq. (5) is well-defined, we require $\|\boldsymbol{A}(\boldsymbol{x}) + t\boldsymbol{\delta}(\boldsymbol{x})\| > 0$ for every $t \in [0,1]$ and $\boldsymbol{x} \in \boldsymbol{U}$. Applying the triangle inequality gives:

$$\|\boldsymbol{A}(\boldsymbol{x})+t\boldsymbol{\delta}(\boldsymbol{x})\| \geq |\|\boldsymbol{A}(\boldsymbol{x})\|-t\|\boldsymbol{\delta}(\boldsymbol{x})\|| = |1-t\|\boldsymbol{\delta}(\boldsymbol{x})\|| \tag{6}$$

so, if $\|\boldsymbol{\delta}(\boldsymbol{x})\| < 1$ we have $1-t\|\boldsymbol{\delta}(\boldsymbol{x})\| \geq 1-\|\boldsymbol{\delta}(\boldsymbol{x})\| > 0$. Hence a sufficient condition for the homotopy $H$ to be well defined is:

$$\|\boldsymbol{\delta}(\boldsymbol{x})\| < 1 \text{ for all } \boldsymbol{x} \in \boldsymbol{U} \tag{7}$$

Notice that if $\|\boldsymbol{A}'(\boldsymbol{x})-\boldsymbol{A}(\boldsymbol{x})\| < 1$ everywhere, this condition is certainly true as we may take $\boldsymbol{\delta}(\boldsymbol{x}) = \boldsymbol{A}'(\boldsymbol{x})-\boldsymbol{A}(\boldsymbol{x})$. Geometrically, this corresponds to $\boldsymbol{A}'-\boldsymbol{A}$ differs by no more than 60° everywhere (see Supplementary Note 4). Using this as a criterion, we can theoretically estimate the magnitude of noise under which the topological number of a field remains unchanged. Note that this practical engineering rule overcomes limitations in experimental measurements, since it is never the case that imperfections in measurements will result in errors as large as 60° (most calibrated Stokes polarimeters are accurate to the order of 1% or less[69]).

*Topological resilience of AGB skyrmions against Gaussian noise*

In this work, AGB skyrmions are formed by cascading three LC-SLMs, resulting in an axis geometry field $\boldsymbol{A}=\boldsymbol{A}(\theta_1,\theta_2,\theta_3)$, where $\theta_1(r,\theta)$, $\theta_2(r,\theta)$, $\theta_3(r,\theta)$ denote the phase patterns applied to each LC-SLM. In practical scenarios, external perturbations can be modeled by altering the phase patterns on the SLMs. To demonstrate the topological stability of AGB skyrmions under perturbations, we introduce Gaussian noise into $\theta_x$, where the random variable $\delta_x$ on a random SLM follows a normal distribution $\delta_x \sim \mathcal{N}(0,\sigma^2)$ Consequently, the perturbed fast-axis vector is modified as follows:

$$\boldsymbol{A}' = \begin{pmatrix} A'_1(r,\theta) \\ A'_2(r,\theta) \\ A'_3(r,\theta) \end{pmatrix} = \frac{1}{\sin\left(\frac{\theta_S}{2}\right)} \begin{pmatrix} \sin\left(\frac{\theta_1+\theta_3+\delta_x Y}{2}\right)\cos\left(\frac{\theta_2+\delta_x(1-Y)}{2}\right) \\ \cos\left(\frac{\theta_3-\theta_1+\delta_x Y}{2}\right)\sin\left(\frac{\theta_2+\delta_x(1-Y)}{2}\right) \\ \sin\left(\frac{\theta_3-\theta_1+\delta_x Y}{2}\right)\sin\left(\frac{\theta_2+\delta_x(1-Y)}{2}\right) \end{pmatrix} \tag{8}$$

where $\theta_s = 0.5\pi$ is the retardance and $Y$ is a random variable which takes value 1 with probability $2/3$ and value 0 with probability $1/3$. As shown in Eq. (9), since the axis geometry is related to the three phase patterns loaded onto the LC-SLMs through trigonometric functions, introducing this random variable leads to highly complex spatial variations in both the radial and azimuthal components of the corresponding field. To understand the impact of noise on the skyrmion number, we performed simulations where we varied the standard deviation of $\delta_x$ from 0 to $\pi$ and computed the resulting axis geometry field and corresponding skyrmion number.

We evaluated the mean skyrmion number $N'_{\mathrm{sk}}$ and its standard deviation $\sigma_{N'_{\mathrm{sk}}}$ at each noise level. These results are shown in Fig. 5. When the standard deviation of the noise $\delta_x$ is small, the condition $\|\boldsymbol{A}'-\boldsymbol{A}\| < 1$ is satisfied (the range of $\sigma$ can be solved as $\sigma \lesssim 0.30\pi$), indicating that the skyrmion is topologically protected. As $\sigma$ increases, not only does $\sigma_{N'_{\mathrm{sk}}}$ grow, but the mean skyrmion number $N'_{\mathrm{sk}}$ progressively decreases towards zero, indicating the gradual destruction of the topological structure. This trend aligns with our experimental observations, where the decline is even more pronounced, possibly due to additional environmental disturbances such as thermal noise that goes beyond the applied Gaussian noise.

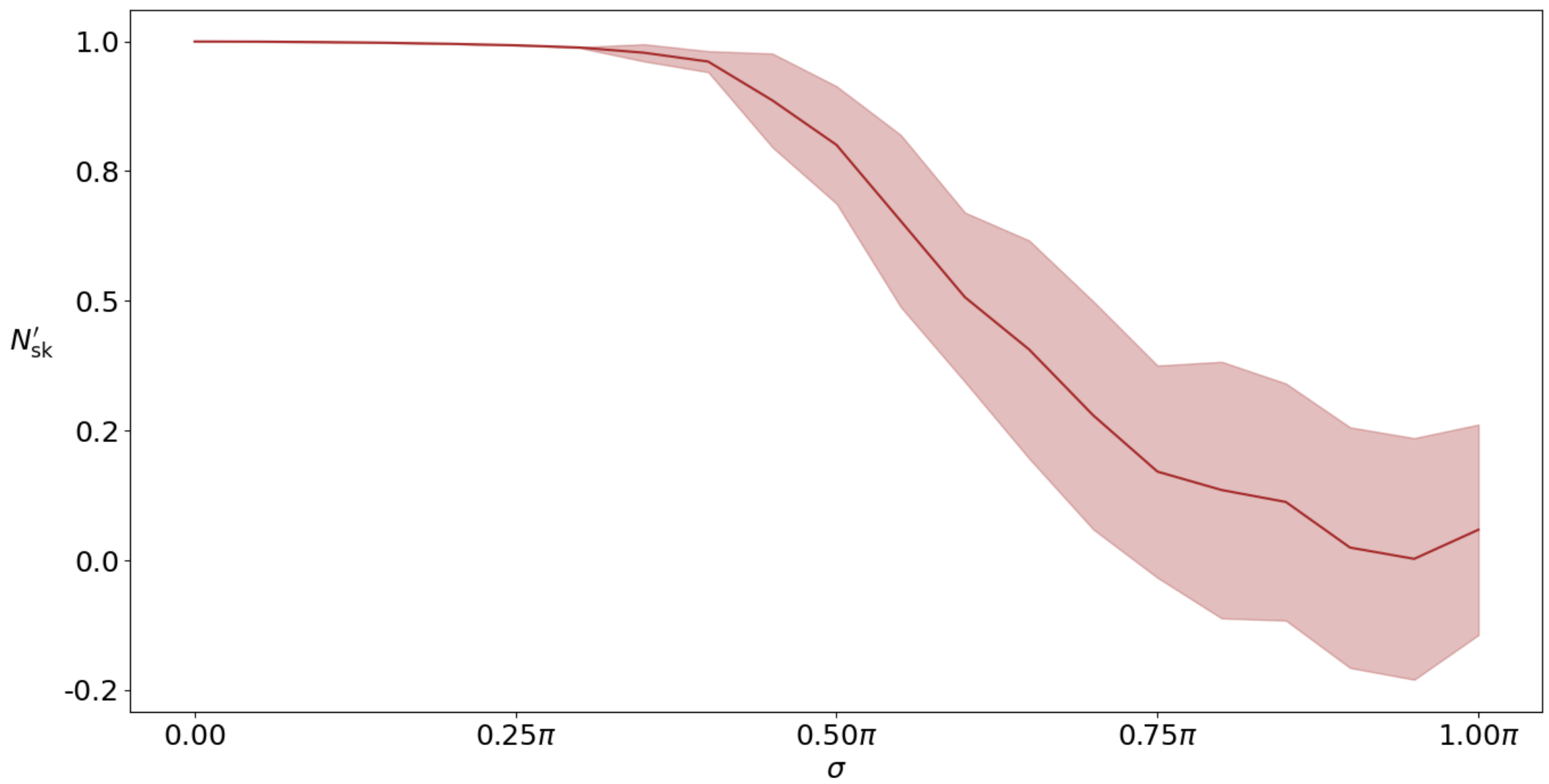

Figure 5 **Simulated skyrmion number under perturbations.** The mean skyrmion number $N'_{\mathrm{sk}}$ (solid red line) and the standard deviation $\sigma_{N'_{\mathrm{sk}}}$ (shaded region) are plotted against the standard deviation $\sigma$ of the added noise $\delta_x$. For $\sigma \leq 0.30\pi$, $N'_{\mathrm{sk}} \approx 1$ and $\sigma_{N'_{\mathrm{sk}}}$ is negligible, confirming the predicted topological protection of the AGB skyrmion. In the range of $0.30\pi < \sigma \leq 0.40\pi$, the AGB skyrmion retains its stability, although the mean skyrmion number decreases slightly. As $\sigma$ increases further, $N'_{\mathrm{sk}}$ progressively decreases, whereas $\sigma_{N'_{\mathrm{sk}}}$ continues to grow, indicating a loss of topological information.

## Acknowledgements

We would like to acknowledge support from the Department of Engineering Science, University of Oxford and the Royal Society University Research Fellowship (URF\R1\241734) (C.H.). The authors thank Prof. Martin J. Booth and Prof. Patrick Salter (University of Oxford), and Prof. Andrew Forbes (University of the Witwatersrand) for valuable discussions and support.

## Author Contributions

All authors reviewed the results and approved the final version of the manuscript.

## Data availability

The data that supports the plots within this paper and other findings of this study are available from C.H. upon reasonable request.

## Conflict of interest

The authors declare no competing interests.

**Additional Information**

Correspondence and request for materials should be addressed to C.H.